\newcommand{\eqb}{\begin{equation}}
\newcommand{\eqe}{\end{equation}}
\newcommand{\dmb}{\begin{displaymath}}
\newcommand{\dme}{\end{displaymath}}
\newcommand{\pad}{\partial}
\newcommand{\ep}{\varepsilon}
\newcommand{\eab}{\begin{eqnarray}}
\newcommand{\eae}{\end{eqnarray}}
\newcommand{\ra}{\rangle}
\newcommand{\la}{\langle}
\newcommand{\e}{\mbox{e}}
\newcommand{\be}{\begin{equation}}
\newcommand{\ee}{\end{equation}}

\newcommand{\eq}{Eq.\ (\ref}
\newcommand{\eqs}{Eqs.\ (\ref}

\documentclass[epj]{svjour}
\usepackage{latexsym}
\usepackage{graphics}
\begin{document}
\title{Thermal QCD sum rules in the $\rho^0$ channel revisited}
\author{R.\ Hofmann \and Th. Gutsche \and Amand Faessler}

\institute{Institut f\"ur Theoretische Physik,
Universit\"at T\"ubingen, 
Auf der Morgenstelle 14, 
72076 T\"ubingen,
Germany}
\date{Received: date / Revised version: date}
%

\abstract{
From the hypothesis that at zero temperature 
the square root of the spectral continuum threshold 
$s_0$ is linearly related to the QCD scale $\Lambda$ we derive in the chiral limit
and for temperatures considerably smaller 
than $\Lambda$ scaling relations for 
the vacuum parts of the Gibbs averaged scalar 
operators contributing 
to the thermal operator product expansion 
of the $\rho^0$ current-current correlator. The scaling with 
$\lambda\equiv \sqrt{s_0(T)/s_0(0)}$, $s_0$ being the $T$-dependent 
perturbative QCD continuum 
threshold in the spectral integral, 
is simple for renormalization group invariant operators, 
and becomes nontrivial for a set of operators which 
mix and scale anomalously under a change of the renormalization point. 
In contrast to previous works on thermal QCD sum rules with this
 approach the gluon condensate exhibits 
a sizable $T$-dependence. The $\rho$\,-meson mass is found to rise 
slowly with temperature 
which coincides with the result found by 
means of a PCAC and current algebra analysis of the $\rho^0$ correlator. 
} 

\maketitle

\section{Introduction}
\label{intro}
QCD sum rules at finite temperature 
have been of intense interest since the 
pioneering work of Bochkarev and Shaposhnikov \cite{BS}. Since thermal field 
theory lacks asymptotically measurable states the 
empirical side, that is the spectral function of the QCD 
sum rule under consideration,
 must be parameterized. There are 
 few guidelines for the parametrization of 
 the spectral function by hadronic resonance and 
 continuum contributions. 
 However, as pointed out by 
several authors \cite{Hatsuda,Leu1,Leu2}, the output 
of the sum rule at finite temperature or 
for nonvanishing chemical potential 
depends 
on the hadronic model 
leading to a specific spectral function. 

Another complication due to finite 
temperature arises from the fact, that temperature 
is defined in a fixed reference frame which hence is singled out. 
Therefore, the Poincar\'e invariance of a field theory at $T=0$ 
is partially broken for finite temperatures. 
Residual O(3) and translational invariance permit 
a wider set of operators to contribute to the 
operator product expansion (OPE) of the corresponding 
correlator. Using a background field method, 
Mallik \cite{Mal} derived an OPE for the time-ordered thermal
correlator of various quark field bilinears, 
where these new operators are included up to mass-dimension four.  
In contrast to the work 
of Hatsuda et. al. \cite{Hatsuda}, 
which relies on the higher twist classification of operators originating from the analysis 
of Deep Inelastic Scattering (DIS) 
and which for nonvanishing spatial momentum components 
allows for contributions of O(3) non-invariant operators 
to the thermal OPE 
\footnote{At first sight this seems to violate rotational symmetry which is retained in 
the heat bath. Invariance with respect to O(3) 
transformations is, however, restored in Ref.\cite{Hatsuda} when performing the 
phase space integrals over pionic matrix elements introduced 
through the definition of the thermal average.},  
a systematic O(3) invariant extension of 
the zero temperature OPE
 is obtained in Ref.\cite{Mal-M}. Thereby the authors 
consider the mixing of the new operators 
under a change of the renormalization scale. However, using the background field method, 
the Wilson-coefficients for the radiative corrections of mass-dimension 
six could not be calculated, 
and hence have been omitted in Ref. \cite{Mal-M}. These 
contributions are important, since they distinguish the vector from the axial vector 
channel and cancel a large part of the terms with 
mass-dimension four, yielding the experimental  
value of the $\rho$\,-meson mass at $T=0$ with good stability \cite{Shuryak}.

In Refs. \cite{BS,Hatsuda,Mal-M} 
the Gibbs average of the OPE 
 is assumed to be saturated by the vacuum and the dilute 
 thermal pion gas contributions. Thereby, both the vacuum and the pion states 
 are taken to be temperature
 independent. 
 This leads to a $T$-independent 
 gluon condensate \cite{Hatsuda}, 
 which is in contrast to lattice measurements \cite{Mil} 
 and to results obtained in effective 
 meson models with scalar glueball fields \cite{Mei,Heinz}. 
 The main point of this paper is to 
 explore the consequences of a $T$-dependent vacuum 
 for the $T$ evolution of the gluon condensate and the $\rho$\,-meson mass 
 in the chiral limit. To this end, we parametrize the 
 $T$-dependence of the vacuum implicitly through the effective 
 spectral variable $s_0(T)$ which divides 
 the hadronic part from the perturbative 
 QCD domain of the spectral function. This ansatz can be justified 
 for renormalization scales of the order of 1
 GeV and for temperatures considerably less than the 
 fundamental QCD scale $\Lambda$ (see Appendix A).
 
 The paper is organized as follows: In section II we list the basic 
 results for thermal dispersion relations and for the spectral function in the $\rho^0$-meson channel  
 as already established in Refs. \cite{Hatsuda} and \cite{Mal-M}. 
 Section III contains the thermal OPE for the invariant longitudinal amplitude of the
  $\rho^0$-meson correlator, 
  and we briefly discuss the behavior of the nonscalar contributions of mass-dimension four 
  under renormalization. 
  The treatment of thermal operator averages is performed in section IV. With an
   implicit dependence on temperature, we derive a scaling relation with respect to 
   $\lambda\equiv\sqrt{s_0(T)/s_0(0)}$ for $T$-dependent 
   vacuum averages of scalar operators. This relation is easily 
  implemented for renormalization group invariants
  (RGI), and becomes quite 
  involved if one has to regard a set of operators 
  which mix and scale anomalously under a change of the
   renormalization point. 
   In section V we write out the Borel transformed 
   sum rule and, by performing a logarithmic derivative 
   with respect to the inverse squared Borel mass, 
   obtain a sum rule for the $\rho$\,-meson mass which is the ratio of two moments.  
   A numerical evaluation of the thermal sum rule is performed in 
   section VI. Section VII summarizes and compares the 
   results with those of previous approaches.

\section{Thermal Dispersion relations, kinematic Invariants, and Spectral Function}
\label{sec:1}

We consider the thermal correlator of the time-ordered (T)
product of two vector currents in the $\rho^0$-channel 
\eqb
\label{thcor}
T_{\mu\nu}(q,T)=Z^{-1}i\int d^4x\ \e^{iq x}\mbox{Tr}\ \e^{-\beta H}\mbox{T}
 j_\mu(x)j_\nu(0)\ ,
\eqe
where 
\eqb
\label{current}
j_\mu(x)=\frac{1}{2}(\bar{u}\gamma_\mu u-\bar{d}\gamma_\mu d)\ 
\ \ \mbox{and}\  Z=\mbox{Tr}\ \e^{-\beta H}\ .
\eqe
Here $H$ denotes the QCD Hamiltonian, 
and $\beta$ stands for the inverse of the temperature $T$.
 A sum rule for this correlator
can be derived as \cite{Mal-M}
\eab
\label{sr}
T_{\mu\nu}(Q_0,|\vec{q}|,T)&=&
\frac{1}{\pi}\int_0^{\infty}{dq_0^\prime}^2\ \frac{\mbox{Im}T_{\mu\nu}(q^\prime_0,|\vec{q}|)}
{{q_0^\prime}^2+Q_0^2}\ \mbox{tanh}(\beta q_0^\prime/2)\ ,\nonumber\\ 
 Q_0^2&\equiv&-q_0^2\ .  
\eae
The frame of reference where 
temperature is defined moves with four-velocity $u_\mu$. With this additional covariant 
one can, for example, define the Lorentz scalars $\omega\equiv u_\mu q^\mu$ and 
$\bar{q}\equiv\sqrt{\omega^2-q^2}$. Imposing current 
conservation and symmetry under exchange of $\mu$ and $\nu$, the correlator of Eq.\ (\ref{sr}) 
can be decomposed as \cite{Hatsuda,Mal-M}
\eqb
T_{\mu\nu}(q,T)=Q_{\mu\nu}T_l(q^2,\omega,T)+P_{\mu\nu}T_t(q^2,\omega,T)\ ,
\eqe
where $(g_{\mu\nu})=\mbox{diag}(1,-1,-1,-1)$, and the tensors $P_{\mu\nu}$, 
$Q_{\mu\nu}$ are given by
\eab
P_{\mu\nu}&=&-g_{\mu\nu}+\frac{q_\mu q_\nu}{q^2}-\frac{q^2}{\bar{q}^2}\tilde{u}_\mu\tilde{u}_\nu
\nonumber\\ 
Q_{\mu\nu}&=&\frac{q^4}{\bar{q}^2}\tilde{u}_\mu\tilde{u}_\nu\ ,\ 
\ \ \ \tilde{u}_\mu\equiv u_\mu-\omega\frac{q_\mu}{q^2}ß .
\eae

By evaluating $\Pi_1\equiv u^\mu u^\nu T_{\mu\nu}$ and 
$\Pi_2\equiv T^\mu_\mu$ in the rest frame of the heat bath ($u_\mu=(1,0,0,0)$),  
one can solve for the invariant amplitudes $T_l$ and $T_t$ \cite{Mal-M} as 
\eqb
\label{T_l}
T_l=\frac{1}{\bar{q}^2}\Pi_2\ ,\ \ \ \ 
T_t=-\frac{1}{2}\left(\Pi_1+\frac{q^2}{\bar{q}^2}\Pi_2\right)\ , 
\eqe
and in the limit $\vec{q}\to 0$ one obtains
\eqb
T_t(q_0,\vec{q}=0)=q_0^2\ T_l(q_0,\vec{q}=0)\ .
\eqe

Since the sum rule of \eq{sr}) holds for each component of ($T_{\mu\nu}$), 
it also holds for $\Pi_1$ and $\Pi_2$. 
With \eq{T_l}) one obtains sum rules for 
the invariant amplitudes $T_l$ and $T_t$. For example,  
\eab
\label{disp_Tl}
T_l(q_0^2,\bar{q},T)&=&\int_0^\infty d{q'^0}^2 
\frac{N_l({q'}^0,\bar{q},T)}{{q'^0}^2 +{Q^0}^2}\ ,\ \mbox{with} \nonumber\\ 
N_l({q'}^0,\bar{q},T)
&\equiv& \pi^{-1} \mbox{Im} T_l\  \mbox{tanh}(\beta {q'^0}/2)\ .
\eae

As already indicated in the introduction, thermal field theory is 
handicapped by the lack of asymptotically 
measurable states, and therefore the spectral 
function (defined as the numerator of the 
integrand of the right-hand side of Eq.\ (\ref{disp_Tl}), 
that is $N_l({q'}^0,\bar{q},T)$),  
must be modelled. In the following $s_0(T)$ denotes 
a temperature dependent threshold that divides 
the hadronic part of the spectrum 
from the perturbatively accessible QCD domain. 
It is suggested \cite{BS,Mal-M}, 
that in the hadronic region of integration, $0\le{q'^0}^2\le s_0$,
the correlator is saturated by the $\rho^0$-resonance and 
the two-pion continuum. Thereby, the pions are
assumed to be noninteracting. The hadronic contributions $j_\mu^{\rho^0}$ and $j_\mu^\pi$ 
to the current $j_\mu$ of \eq{current}) 
are obtained by using the field-current-identity
\eqb
j^{\rho^0}_\mu=m_\rho f_\rho\rho^0_\mu\ ,\ \ 
\ \mbox{where}\ \  \ f_\rho(T=0)=153.5 \ \mbox{MeV}\ ,
\eqe
and by appealing to the SU(2) flavor symmetry structure resulting in  
\eqb
j_\mu^\pi=\ep^{3bc}\pi^b\pad_\mu\pi^c\ .
\eqe
Thereby, the pionic current is obtained by constructing the Noether current of the 
SU(2)$_V$ symmetry of a low-energy chiral Goldstone-field theory 
which to lowest order in the derivatives contains noninteracting fields. Since we only consider 
a dilute pion gas (one-pion states in Gibbs average) this free 
pion theory is relevant for our purposes.   

Employing the noninteracting, 
finite temperature (real-time) $\rho^0$-meson and pion propagators \cite{Weld,Kapust,Landsman}
for the calculation of the unitarity cuts for 
the tree diagram of the $\rho^0$ contribution and for the thermal pion loop, 
the hadronic part of the spectral function 
has been calculated in Ref.\cite{Mal-M}. 
In contrast to Ref.\cite{Mal-M} and following Ref.\cite{Hatsuda}
 we approximate the imaginary part of the 
correlator above the threshold $s_0(T)$ 
by perturbative QCD (pQCD). For this perturbative piece thermal contributions
are omitted since the fermionic distribution function $n_F$ remains small 
for the relevant temperature and energy range \cite{Hatsuda}. In the limit 
$\vec{q}\to 0$ and after a Borel transformation in $Q_0$ with 
Borel mass $M$ \cite{SVZ} the spectral side of the sum rule of \eq{disp_Tl})
 reads 
\eab
\label{emp}
T_l(M,T)&=&\frac{1}{\pi M^2}\int_0^\infty ds\
 \mbox{Im}T_l(s,T)\ \e^{-s/M^2}\mbox{tanh}(\sqrt{s}/(2T))\nonumber\\ 
 &\equiv&\frac{1}{M^2}\int_0^\infty ds\ \e^{-s/M^2}\ \rho(s,T)\nonumber\\ 
&=&\frac{1}{M^2}\left(f_\rho^2(T)\e^{-m_\rho^2(T)/M^2}+
J_0^{\pi\pi}+J_T^{\pi\pi}+J_0^{qq}\right)\nonumber\\ 
& &\mbox{with} \ s\equiv {q'^0}^2\ ,
\eae
and  
\eab
\label{cont}
J_0^{\pi\pi}&=&\frac{1}{48\pi^2}\int_{4m_\pi^2}^{s_0(T)}ds\ \e^{-s/M^2} v^3\ ,\  
J_0^{qq}=\frac{1}{8\pi^2}\left(1+\frac{\alpha_s(\mu)}{\pi}\right)\nonumber\\ 
J_T^{\pi\pi}&=&\int_{s_0(T)}^{\infty}ds\ \e^{-s/M^2} 
\frac{1}{24\pi^2}\int_{4m_\pi^2}^{s_0}ds\nonumber\\ 
& &(\e^{-s/M^2} v^3+v(3-v^2)/2)\ n_B(\sqrt{s}/2,T)\ ,
\eae
where $\mu=1\ \mbox{GeV}$. Here the function $v$ is defined as 
$v(s,m_\pi)=\sqrt{1-4m_\pi^2/s}$, $n_{B}$ denotes 
the Bose distribution, and $J_{0(T)}^{\pi\pi}$, $J_{0}^{qq}$ are the
vacuum (thermal) parts of the 
spectral integrals due to the $\pi\pi$ and pQCD continua, respectively.

\label{sec:2}
\section{Thermal Operator Product Expansion}

In this work we use a thermal OPE for the invariant amplitude $T_l$ 
which combines the results of Ref.\cite{Mal-M} 
 and Ref.\cite{Hatsuda}. In Ref.\cite{Mal-M} 
 the expansion is 
only carried out up to operators of mass-dimension four. The 
nonscalar O(3) invariant contributions 
are expressed in terms of diagonal combinations with respect to the 
anomalous mixing matrix, resulting in a renormalization group invariant (RGI)  
(that is the total energy density) and a renormalization group non-invariant (RGNI) contribution. 
The drawback of the OPE of Ref.\cite{Mal-M} lies in the fact, 
that mass-dimension six contributions are omitted. These terms 
are important since they cancel a large part of the 
nonperturbative correction of mass-dimension four yielding at $T=0$ 
the experimentally measured $\rho$\,-meson mass. 
In addition, it is the mass-dimension six part of 
the OPE that distinguishes at $T=0$ the vector from the axial
vector channel \cite{Shuryak}. Therefore, we use together with 
the standard $T=0$ scalar operators the results 
of Ref.\cite{Hatsuda} for the nonscalar 
mass-dimension six contribution. In the chiral limit, for  
$\vec{q}\to 0$, and for two light quark flavors ($u$ and $d$)
\footnote{In order to 
compare our results 
with the lattice data of Ref. \cite{Mil} for two quark flavors 
we constrain ourselves 
to QCD with two light quark flavors throughout the paper.}, we then
obtain the following expansion for the invariant $T_l$ 
\eab
\label{OPETl}
& &T_l({q^0}^2,\vec{q}=0,T)=
\nonumber\\ 
& &-\frac{1}{8\pi^2}
\mbox{ln}\left(\frac{Q_0^2}{\mu^2}\right)(1+
\frac{\alpha_s(\mu^2)}{\pi})+
\frac{1}{Q_0^4}\left\{\frac{1}{24}\la\frac{\alpha_s}{
\pi}F^a_{\mu\nu}{F^{\mu\nu}}_a\ra_T(\mu)+\right.\nonumber\\ 
& &\left.\frac{2}{11}\left[\la \theta_{00}\ra_T(\mu)+
\left(\frac{\alpha_s(\mu^2)}
{\alpha_s(Q_0^2)}\right)^{\delta/b}
\la\frac{16}{3}\theta_{00}^f-\theta_{00}^g\ra_T(\mu)\right]\right\}-\nonumber\\ 
& &\frac{\pi}{2 Q_0^6} 
\la \alpha_s (\bar{u}\gamma_\mu\gamma_5 t^a u-
\bar{d}\gamma_\mu\gamma_5 t^a d)^2\ra_T(Q_0)-\nonumber\\ 
& &\frac{\pi}{9 Q_0^6} \la \alpha_s 
(\bar{u}\gamma_\mu t^a u+\bar{d}\gamma_\mu t^a d)^2\ra_T(Q_0)+\nonumber\\ 
& &\frac{8\pi i}{3 Q_0^6} \la \bar{u}\gamma_0D_0D_0D_0 u+
\bar{d}\gamma_0D_0D_0D_0 d\ra_T(\mu)  ,
\eae
where the nonscalar operators are understood 
to be symmetrized and to be made traceless with respect to the Lorentz
indices. The generators $t^a$ of color 
SU(3) in the fundamental representation are normalized to Tr$t^at^b=2\ \delta^{ab}$. 
There are, in principle, also O(3) mixed quark-gluon operators of twist 4. 
However, their there is no experimental clue about their matrix elements 
as pointed out in Ref.\,\cite{Hatsuda}. There it was also indicated 
that bag model estimates of the nucleon matrix elements of 
these operators yield very small values and that the treatment of pions in the bag model is 
rather questionable due to their collective nature. Following Ref.\,\cite{Hatsuda} 
we simply omit these operators to obtain Eq.\,(\ref{OPETl}).    

The Gibbs average in \eq{OPETl}) is approximated by the 
vacuum and the dilute pion gas contributions in Ref. \cite{Hatsuda}. 
 As the authors point out the scalar 
part of the Gibbs averaged OPE of \eq{OPETl}) then 
has the same structure as the result 
for the vector correlator obtained in Ref. \cite{Dey}. 
Based on PCAC, current algebra, and the LSZ reduction formula the authors of Ref. \cite{Dey} 
find to order $T^2$ a mixing of the correlator of the vector with that of the axial 
vector channel due to finite temperature. 

To one loop and for two quark flavors 
the constants $b$ and $\delta$ are given by \cite{Peskin}
\eqb
\label{beta}
\delta=-\frac{2}{3}(\frac{16}{3}+2),\ \ \ b=11-2\ \frac{2}{3}\ .
\eqe
At $\mu=1$ GeV we take $\alpha_s(\mu^2)=0.36$ \cite{Leu1}. 
Note, that in the case of mass-dimension four the operators 
a priori renormalized at $Q^0$ 
are already expressed by operators evaluated at $\mu=1$ GeV. 
Only for $\la\frac{16}{3}\theta_{00}^f-\theta_{00}^g\ra_T$ 
does this process of rescaling 
generate a renormalization group logarithm due to the nonvanishing 
anomalous dimension $\delta$ of this operator.  

The operator $\frac{\alpha_s}{\pi}F^a_{\mu\nu}{F^{\mu\nu}}_a$ 
is an RGI. In the chiral limit, the fermionic and gluonic parts 
of the 00-component of the traceless 
energy-momentum tensor ($\theta_{\mu\nu}$) 
of one quark flavor QCD \cite{Mal-M,Tarrach} are
\eab
\label{O3}
\theta^f_{00}&=&i\bar{q}\gamma_0D_0 q\ ,\nonumber\\ 
\theta^g_{00}&=&-F^a_{0\lambda}{F^{\lambda}_{\,0}}_a+\frac{1}{4}
g_{00}\ F^a_{\kappa\lambda}{F^{\kappa\lambda}}_a\ ,\nonumber\\ 
D_0&\equiv&\pad_0-ig A^a_0 \frac{t_a}{2}\ ,
\eae
with $\theta_{00}$ given by 
\eqb
\theta_{00}=\theta^f_{00}+\theta^g_{00}\ .
\eqe
Since $\theta_{\mu\nu}$ is a conserved quantity 
the thermal average of $\theta_{00}$ is also an RGI.
 The expression for the covariant derivative $D_0$ in Eq.\ (\ref{O3}) 
contains the zeroth component of the gauge potential $A^a_0$ 
and the color SU(3) generators $t_a$ in the fundamental 
representation normalized to Tr$t_a t_b=2\delta_{ab}$.   

In \eq{OPETl}) both mass-dimension six scalar operators are RGNI, 
and we will consider their anomalous scaling and mixing in the next section. As for 
the O(3), non-scalar mass-dimension six contribution to \eq{OPETl}) we neglect the anomalous
scaling of the corresponding operator \cite{Hatsuda}. It is known to  
belong to a set of operators which mix under renormalization. This set also 
contains mixed quark-gluon contributions, and following Ref.\cite{Hatsuda} 
we will omit them. 

\section{Gibbs averages}
\label{sec:3}
\subsection{Scalar operators}
\label{subsec:3.1}

As was suggested by several authors \cite{BS,Hatsuda,Mal-M} the Gibbs averages 
of the scalar operators of Eq.\ (\ref{OPETl}) 
can be saturated by the vacuum and the one pion contributions.  
Thereby, both the vacuum and the pion states were assumed to exhibit no 
temperature dependence \cite{Hatsuda}. In the chiral limit the 
only quantities entering the sum rule, which can 
potentially describe thermal pion properties, 
are the pion decay constant $f_\pi$ and the pion matrix 
elements of twist two operators. As for the former the $T$ 
dependence has been obtained in the imaginary time formulation 
of thermal chiral perturbation theory in Ref.\,\cite{Gasser}. 
There, the result for $f_\pi(T)$ is a notably decreasing 
function of temperature for $T\ge f_\pi$. On the contrary, 
the lattice simulation of 
Ref.\cite{BoydGupta} obtains a nearly 
$T$ independent pion decay 
constant up to the critical temperature which 
is understood as 
an artefact due to the large pion mass 
($\sim 400$ MeV) used. 
As for the $T$-dependence 
of the twist two pion averages 
no information is available so far, and we have to 
use the pion-in-vacuum parton distributions to 
estimate the corresponding matrix elements. 
In accord with previous sum rule investigations at finite 
temperature \cite{Mal-M,Hatsuda,BS} and for consistency 
we then have no choice but to assume a 
$T$-independent pion decay constant. 

The motivation for a $T$-dependent vacuum 
part in the Gibbs average of scalar operators 
emerges from a comparison of lattice data for the 
gluon condensate \cite{Mil} with the result of Ref.\cite{Hatsuda}.      
The former approach indicates a sizable decrease above a  critical 
temperature of about $T_c=140$ MeV, whereas in the 
latter case the gluon condensate exhibits practically no 
temperature dependence even when leaving the chiral limit 
(less than 0.5\% decrease at $T=200$ MeV)\footnote{
On the lattice and at $T=0$ the ground state average of the local two 
gluon operator $F^a_{\mu\nu}(0)F_a^{\mu\nu}(0)$ 
in contrast to the gluon condensate used in an OPE 
contains also perturbative contributions 
which must be subtracted for a direct comparison. 
However, at finite temperature the perturbative 
part, which contributes to the partition function 
for momenta larger than, say, 1.5 GeV, is for $T<200$ MeV severely Boltzmann suppressed 
and hence hardly influences the $T$ dependence of the thermal average.}. 
On a more phenomenological level, 
the $T$-dependence of the gluon condensate has 
been obtained from the effective potential of theories
based on the non-linear $\sigma$-model, where the 
mesons are coupled to a scalar glueball field to 
mimic the breaking of scale invariance by the QCD vacuum \cite{Mei,Heinz}. 
Depending on the normalization conditions used for the Bag constant and the glueball mass, 
in these calculations the $T$-evolution of the gluon condensate exhibits a strong decrease 
above critical temperatures ranging from $T_c=140-400$ MeV. 

The scalar operators appearing in the OPE of the thermal current correlator 
read
\eab
\label{scalO}
\mbox{dim}\, 4:
\ \ \ &&{\cal O}^s_4=\frac{\alpha_s}{\pi}\ F^a_{\mu\nu}F_a^{\mu\nu}\nonumber\\ 
\mbox{dim}\, 6:
\ \ \ &&{\cal O}^s_{6,1}=\pi\alpha_s\left(\bar{u}\gamma_\mu\gamma_5t^a u-
\bar{d}\gamma_\mu\gamma_5t^a d\right)^2\ , \nonumber\\ 
&&{\cal O}^s_{6,2}=\pi\alpha_s\left(\bar{u}\gamma_\mu t^a u+
\bar{d}\gamma_\mu t^a d\right)^2\ .
\eae
Hereby, ${\cal O}^s_4$ 
is an RGI, and ${\cal O}^s_{6,1}$, 
${\cal O}^s_{6,2}$ can be expanded into a 
basis of scalar four-quark operators $P_1, P_2, ..., P_6$ \cite{SVZ} with 
\eab
\label{Ps}
P_1&=&\bar{\psi}_L\gamma_\mu\psi_L\bar{\psi}_R\gamma_\mu\psi_R\nonumber\\
P_2&=&\bar{\psi}_L\gamma_\mu t^a\psi_L\bar{\psi}_R\gamma_\mu t^a\psi_R
\nonumber\\
P_3&=&\bar{\psi}_L\gamma_\mu\psi_L\bar{\psi}_L\gamma_\mu\psi_L+(L\to R)
\nonumber\\        
P_4&=&\bar{\psi}_L\gamma_\mu t^a\psi_L
\bar{\psi}_L\gamma_\mu t^a\psi_L+(L\to R)\nonumber\\  
P_5&=&\bar{\psi}_L\gamma_\mu\lambda^bt^a\psi_L
\bar{\psi}_R\gamma_\mu\lambda^bt^a\psi_R\nonumber\\
P_6&=&\bar{\psi}_L\gamma_\mu\lambda^b\psi_L
\bar{\psi}_R\gamma_\mu\lambda^b\psi_R\ ,\nonumber\\ 
\psi_{L(R)}&=&(u_{L(R)},d_{L(R)})^T\ ,
\eae
where $t^a$ and $\lambda^b$ are the color SU(3) Gell-Mann and flavor SU(2) Pauli
 matrices, respectively. They are normalized to 
$\mbox{Tr}{t^at^b}=\mbox{Tr}{\lambda^a\lambda^b}=2\delta^{ab}$. 
The left-(right-) handed spinors are given by
\eqb
\psi_{L(R)}=\frac{1}{2}(1\pm\gamma_5)\psi\ .
\eqe
Using the relation 
\eqb
\tau^c_{ij}\tau^c_{mn}=
2(\delta_{in}\delta_{jm}-\frac{1}{N}\delta_{ij}\delta_{mn})
\eqe
for the SU(N) 
generator matrices $\tau^c$ ($\mbox{Tr}{\tau^a\tau^b}=2\delta^{ab}$)
and applying Fierz transformations, one 
obtains the following decomposition of the flavor singlet parts\footnote{
Notationally not quite correct we also refer to them as
 ${\cal O}^s_{6,1}, {\cal O}^s_{6,2}$.} (for a derivation see the Appendix B) 
\eab
\label{O6dec}
{\cal O}^s_{6,1}&=&\pi\alpha_s\frac{1}{3}
\left(-\frac{5}{3}P_4+\frac{32}{9}P_3-2\ P_5\right)\nonumber\\
{\cal O}^s_{6,2}&=&\pi\alpha_s\left(P_4+2P_2\right)\ .
\eae
The sets of operators $P_1,...,P_4$ and $P_5,P_6$ mix independently
 under renormalization with the respective one-loop 
mixing matrices ${\bf \delta}$ and ${\bf \tilde{\delta}}$  \cite{Barfoot}
\eqb
\label{Mix}
{\bf \delta}=\left(\begin{array}{cccc}
0&3/2&0&0\\ 
16/3&17/3&0&-2/3\\ 
0&-2/3&0&-11/6\\ 
0&-20/9&-16/3&8/9
\end{array}\right)\ ,\ \ 
{\bf \tilde{\delta}}=\left(\begin{array}{cc}
7&16/3\\ 
3/2&0  
\end{array}\right)\ .
\eqe
The eigenvalues (proportional to the anomalous dimensions of the diagonal combinations) read
\eab
\label{delta}
\delta_1&=&7.043\ ,\ \delta_2=3.501\ ,\ \delta_3=-2.891\ ,
\delta_4=-1.097\ ,\nonumber\\ \tilde{\delta}_5&=&8\ , \mbox{and}\ \tilde{\delta}_6=-1\ .
\eae

In Appendix A we argue that for temperatures considerably 
smaller than the fundamental QCD scale 
$\Lambda$ ($\Lambda\approx 200$ MeV \cite{SVZ,Peskin}), in the chiral limit, and for a 
renormalization scale $Q_0$ of the order of 1 GeV 
it should be possible to describe the $T$ 
dependence of the thermal average of a scalar 
operator implicitely through 
that of the $T$-dependent spectral continuum threshold $s_0(T)$. 
Meeting the above premises, we can derive a scaling relation 
with respect to $\lambda\equiv\sqrt{s_0(T)/s_0(0)}$ for the 
vacuum average of a given scalar and diagonal operator ${\cal O}$ with mass-dimension $d$ and 
anomalous dimension one-loop coefficient $\delta_O$. 
In the spirit of a random phase approximation, where correlations 
between particle-hole excitations in the ground state 
are included, we set up the QCD vacuum state as an expansion 
in terms of $k$-particle (off-shell) 
quark-antiquark and gluonic fluctuations with vacuum quantum numbers, that is   
\eab
|0\ra&=&\sum_k |k\ra, \ \mbox{with}\nonumber\\ 
|k\ra&=&\sum_{i_1,i_2,...,i_k}\int d^4p_1\int d^4p_2\cdots \int d^4p_k \nonumber\\
& &C_{i_1,i_2,...,i_k}(p_1,p_2,...,p_k)\ |p_1;i_1\ra\cdots|p_k;i_k\ra\ .\nonumber\\ 
\eae
Hereby, $i_j$ is a collective index labelling the particle 
species (fermion or boson), 
spatial quantum numbers, flavor (if fermionic), and color. The $C_{i_1,i_2,...,i_k}$ denote the
corresponding expansion coeffcients. Then the vacuum average  
of ${\cal O}$ has a representation of the following form
\eab
\label{repcondh}
\la0|{\cal O}|0\ra(Q_0)&=&\sum_k\sum_{i_1,i_2,...,i_k}
\int d^4p_1\int d^4p_2\cdots \int d^4p_k \nonumber\\  
& &f_{i_1,i_2,...,i_k}(p_1,p_2,...,p_k;Q_0)\ ,\nonumber\\ 
\eae
where the functions $f_{i_1,i_2,...,i_k}$ have mass-dimension $d-4k$, 
and $Q_0$ denotes the scale at which
the operator $\cal O$ is renormalized. 
By asymptotic freedom, the integrand of 
\eq{repcondh}) will be strongly suppressed, if the momenta $p_1,...,p_k$ are hard. 
The finite value of $\la0|{\cal O}|0\ra(Q_0)$ has its origin in the strong 
dynamics of soft fluctuations in the vacuum \cite{SVZ}. A criterion distinguishing   
soft from hard fluctuations should roughly be given by the scale $s_0$. For sizable contributions 
to the integral of \eq{repcondh}) we assume   
that the time- and space-like virtuality $p_j^2$ of each of the momenta 
$p_1,...,p_k$ be less than $s_0$ and, in addition, that $(p^0_j)^2$ be less then $s_0$ 
(recall that through the
presence of the heat bath $(p^0_j)^2$ 
is formally raised to a Lorentz scalar). With the above premises and the results of Appendix A 
we may assume that temperature effectively acts on the vacuum only
implicitely through $s_0$ and that there is a linear relation between the $T$ 
dependent QCD scale $\Lambda_T$ and $\sqrt{s_0(T)}$.  
Thus there are only two independent scales to be considered: $s_0(T)$ and $Q_0$. 
Performing the integrations over the spatial 
components of the momenta in \eq{repcondh}) we obtain
\eab
\label{repcond}
&&\la0|{\cal O}|0\ra(Q_0)=\nonumber\\ 
&&\sum_k\sum_{i_1,i_2,...,i_k}\int_{\tiny{-\sqrt{s_0(0)}}}^{\tiny{\sqrt{s_0(0)}}} 
dp^0_1\cdots \int_{\tiny{-\sqrt{s_0(0)}}}^{\tiny{\sqrt{s_0(0)}}} dp^0_k\nonumber\\   
& &h_{i_1,i_2,...,i_k}(p^0_1,p^0_2,...,p^0_k)\times \nonumber\\ 
& &g_{i_1,i_2,...,i_k}
\left(\left\{\frac{p_1^0}{p^0_k},\cdots,\frac{p_{k-1}^0}{p^0_k}\right\},
\left\{\frac{p_1^0}{Q_0},\frac{p_1^0}{\sqrt{s_0(0)}}\right\},\cdots,\right.\nonumber\\  
& &\left.\left\{\frac{p_k^0}{Q_0},\frac{p_k^0}{\sqrt{s_0(0)}}\right\};
\frac{Q_0}{\sqrt{s_0(0)}}\right)\ .\nonumber\\ 
\eae
In \eq{repcond}) the functions $h_{i_1,i_2,...,i_k}$ are 
homogeneous functions of $p^0_i,\ (i=1,\cdots,k)$ with mass-dimension $d-k$, 
and $g_{i_1,i_2,...,i_k}$ are
dimensionless functions of their dimensionless arguments. 
The $T$-dependent vacuum average
(not to be confused with the Gibbs average) is then given as 
\eab
\label{deftvac}
&&\la0|{\cal O}|0\ra_T(Q_0)=\nonumber\\ 
&&\sum_k\sum_{i_1,i_2,...,i_k}\int_{\tiny{-\sqrt{s_0(T)}}}^{\tiny{\sqrt{s_0(T)}}} 
dp^0_1\cdots \int_{\tiny{-\sqrt{s_0(T)}}}^{\tiny{\sqrt{s_0(T)}}} dp^0_k\nonumber\\ 
& &h_{i_1,i_2,...,i_k}
(p^0_1,p^0_2,...,p^0_k)\times \nonumber\\ 
& &g_{i_1,i_2,...,i_k}
\left(\left\{\frac{p_1^0}{p^0_k},\cdots,\frac{p_{k-1}^0}{p^0_k}\right\}, 
\left\{\frac{p_1^0}{Q_0},\frac{p_1^0}{\sqrt{s_0(T)}}\right\},\cdots,\right.\nonumber\\  
& &\left.\left\{\frac{p_k^0}{Q_0},\frac{p_k^0}{\sqrt{s_0(T)}}\right\};
\frac{Q_0}{\sqrt{s_0(T)}}\right)\ .\nonumber\\ 
\eae
With
\eqb
\lambda=\sqrt{\frac{s_0(T)}{s_0(0)}} \ ,
\eqe
and from comparison of \eq{repcond}) and \eq{deftvac}) one easily obtains 
\eqb
\label{scaling}
\la0|{\cal O}|0\ra_T(\lambda Q_0)=
\lambda^d\ \la0|{\cal O}|0\ra
(Q_0)\ .
\eqe
Eq.\ (\ref{scaling}) relates the average of a diagonal operator ${\cal O}$ 
with mass dimension $d$ taken in a temperature destorted 
vacuum and {\em renormalized at $\lambda Q_0$} 
to the $T=0$ vacuum average {\em renormalized at $Q_0$}. 
This scaling relation should embody a good approximation 
for the thermal vacuum average, provided 
that the above premises of massless quarks, $T$ considerably smaller than $\Lambda$, and 
$Q_0$ of the order of 1 GeV are fulfilled.  

For an RGI operator ${\cal O}$ ($\delta_O$=0) the rescaling from $\lambda Q_0$ to $Q_0$ 
is trivial (for example ${\cal O}^s_4$). 
The situation becomes more involved for the scalar 
operators ${\cal O}^s_{6,1}$ and ${\cal O}^s_{6,2}$ of mass-dimension 
six given in \eq{scalO}), and we will focus on them now. 

The perturbative one-loop renormalization group rescaling from $\lambda Q_0$ 
to $Q_0$ for the diagonal operator ${\cal O}$ is given by
\eqb
\la0|{\cal O}|0\ra_T(Q_0)=\left(\frac{\mbox{log}(\lambda Q_0)^2/\Lambda^2}
{\mbox{log}\ Q_0^2/\Lambda^2}\right)^{-(\delta_{\cal O}/b)}\ 
\la0|{\cal O}|0\ra_T(\lambda Q_0)\ ,
\eqe
where $b$ is defined in \eq{beta}).
In the OPE, operators renormalized at 
$Q_0$ are expressed by operators renormalized at a 
common reference point $\mu$. \eq {scaling}) implies that first we 
rescale from $\lambda Q_0$ to $Q_0$ which is accomplished by
\eqb
\la0|{\cal O}|0\ra_T(Q_0)=\lambda^d\left(\frac{\mbox{log}(\lambda Q_0)^2/\Lambda^2}
{\mbox{log}\ Q_0^2/\Lambda^2}\right)^{-(\delta_{\cal O}/b)}\ 
\la0|{\cal O}|0\ra(Q_0)\ ,
\eqe
and then express $\la0|{\cal O}|0\ra(Q_0)$ 
by $\la0|{\cal O}|0\ra(\mu)$ as 
\eab
\label{tscal}
\la0|{\cal O}|0\ra_T(Q_0)&=&\lambda^d\left(\frac{\mbox{log}(\lambda Q_0)^2/\Lambda^2}
{\mbox{log}\ Q_0^2/\Lambda^2}\right)^{-(\delta_{\cal O}/b)}\times\nonumber\\ 
&&\left(\frac{\mbox{log}\ Q_0^2/\Lambda^2}
{\mbox{log}\ \mu^2/\Lambda^2}\right)^{\delta_{\cal O}/b}
\la0|{\cal O}|0\ra(\mu)\ .
\eae
In order to apply the following identity for the Borel transformation ${\bf L}_M$ \cite{SVZ} 
\eab
\label{bor}
&&{\bf L}_M\left[\left(\frac{1}{Q_0^2}\right)^k\left(\frac{1}
{\mbox{ln}(Q_0^2/\Lambda^2)}\right)^\ep\right]
=\nonumber\\ 
& &\frac{1}{\Gamma(k)}\left(\frac{1}{M^2}\right)^k
\left(\frac{1}{\mbox{ln}(M^2/\Lambda^2)}\right)^\ep\times\nonumber\\ 
& &\left[1+
O\left(\frac{1}{\mbox{ln}(M^2/\Lambda^2)}
\right)\right]\ ,
\eae
we expand $\left(\frac{1}{Q_0^2}\right)^k
\la0|{\cal O}|0\ra_T(Q_0)$ about $\lambda^2=1$ up to quadratic order and perform 
afterwards the Borel transformation indicated in \eq {bor}). The result is
\eab
\label{borex}
&&{\bf L}_M\left[\left(\frac{1}{Q_0^2}\right)^k\ \la0|{\cal O}|0\ra_T(Q_0)\right]
\approx\nonumber\\ 
&&\frac{\lambda^{2k}}{\Gamma(k)}\left(\frac{1}{M^2}\right)^k\left\{
\left(\mbox{ln}(M^2/\Lambda^2)\right)^{\delta_{\cal O}/b-1}-\right.\nonumber\\
&&\left.(\delta_{\cal O}/b)
\left(\mbox{ln}(M^2/\Lambda^2)\right)^{(\delta_{\cal O}/b-2)}(\lambda^2-1)+\right.\nonumber\\
&&\left.\frac{\delta_{\cal O}}{2b^2}\left[\delta_{\cal O}+b\left\{1+\mbox{ln}(\mu^2/\Lambda^2)\right\}\right]
\left(\mbox{ln}(M^2/\Lambda^2)\right)^{(\delta_{\cal O}/b-3)}(\lambda^2-1)^2+\right.\nonumber\\ 
&&\left.{O}((\lambda^2-1)^3)\right\}
\left(\mbox{ln}(\mu^2/\Lambda^2)\right)^{(1-\delta_{\cal O}/b)}\la0|{\cal O}|0\ra(\mu)\nonumber\\ 
&\equiv &\frac{\lambda^{2k}}{\Gamma(k)}\left(\frac{1}{M^2}\right)^k\left\{c_0+
c_1(\lambda^2-1)+c_2(\lambda^2-1)^2+\right.\nonumber\\ 
&&\left.{O}((\lambda^2-1)^3)\right\}
\left(\mbox{ln}(\mu^2/\Lambda^2)\right)^{(1-\delta_{\cal O}/b)}\la0|{\cal O}|0\ra(\mu)\ .
\eae
With $M^2\approx 0.75 \ 
\mbox{GeV}^2$ and $\Lambda^2=0.04 \ \mbox{GeV}^2$ the coefficient 
$c_2$ in the expansion of \eq {borex}) is roughly given by $1/2 c_1$. So even for 
a value of $\lambda^2$ as low as 1/2 
we obtain a suppression for the quadratic as compared to the
linear term in \eq {borex}) by a factor 
of four. 

In the case of the two sets $P_1,..., P_4$ and 
$P_5, P_6$ of \eq{Ps}) the mixing under renormalization results in 
\eab
\label{mixla}
\la0|\pi\alpha_s\left(\begin{array}{c}
P_1\\
\vdots\\
P_4
\end{array}\right)|0\ra_T(M)&=&{\bf T} {\bf D} {\bf T}^{-1}
\la0|\pi\alpha_s\left(\begin{array}{c}
P_1\\
\vdots\\
P_4
\end{array}\right)|0\ra(\mu)
\nonumber\\ 
\la0|\pi\alpha_s\left(\begin{array}{c}
P_5\\
P_6
\end{array}\right)|0\ra_T(M)&=&\tilde{{\bf T}} \tilde{{\bf D}} 
\tilde{{\bf T}}^{-1}\la0|\pi\alpha_s\left(\begin{array}{c}
P_1\\
P_2
\end{array}\right)|0\ra(\mu)\ ,\nonumber\\ 
\eae
where the matrices ${\bf T}$ and $\tilde{{\bf T}}$, transforming 
the matrices ${\bf \delta}$ and ${\bf \tilde{\delta}}$ of \eq{Mix}),
respectively, are given by 
\eab
 \label{Transf}
{\bf T}&=&\left(\begin{array}{cccc}
0.1965&-0.0556&0.0496&0.6838\\ 
0.9224&-0.1298&-0.0955&-0.5003\\ 
-0.0008&0.4786&-0.5481&0.3554\\ 
-0.3324&-0.8666&-0.8295&0.3947
\end{array}\right)\ ,\nonumber\\   
\tilde{{\bf T}}&=&\left(\begin{array}{cc}
8/9&-16/27\\ 
1/6&8/9 
\end{array}\right)\ .
\eae
The diagonal matrices 
${\bf D}=\mbox{diag}(D_1,D_2,D_3,D_4)$ and 
$\tilde{{\bf D}}=\mbox{diag}(D_5,D_6)$ 
have matrix elements of the form
\eab
\label{diagel}
&&D_i=\lambda^6\left\{
\left(\mbox{ln}(M^2/\Lambda^2)\right)^{(\delta_i/b-1)}-\right.\nonumber\\ 
& &\left.(\delta_i/b)
\left(\mbox{ln}(M^2/\Lambda^2)\right)^{(\delta_i/b-2)}(\lambda^2-1)\right.+\nonumber\\
& &\left.\frac{\delta_i}{2b^2}\left(\delta_i+b(1+\mbox{ln}(\mu^2/\Lambda^2))\right)
\left(\mbox{ln}(M^2/\Lambda^2)\right)^{(\delta_i/b-3)}\times\right.\nonumber\\ 
& &\left.(\lambda^2-1)^2+{\cal O}((\lambda^2-1)^3)\right\}\nonumber\\ 
& &\left(\mbox{ln}(\mu^2/\Lambda^2)\right)^{(1-\delta_i/b)}\ ,\ \ \ \ \ \ \ \ \ \ \ 
 (i=1,...,6)\ ,
\eae
where $\delta_1,...,\delta_6$ are given in \eq{delta}). 
Note that the one-loop scaling of $\alpha_s$ is also 
included in ${\bf D}$ and $\tilde{{\bf D}}$. When evaluating the sum rule numerically
 we will consider 
a truncation of $D_i$ after terms of zeroth and second order in $(\lambda^2-1)$ 
to test the sensitivity of the $T$-evolution of 
$m_\rho$, $s_0$, and the gluon condensate.  

Making use of the vacuum saturation hypothesis of 
Shifman, Vainshtein, and Zakharov (SVZ) \cite{SVZ}
\eqb
\label{vacsat}
\la0|\bar{\psi}\Gamma_1\psi\bar{\psi}\Gamma_2\psi|0\ra=
N^{-2}[\mbox{Tr}\Gamma_1\mbox{Tr}\Gamma_2-\mbox{Tr}\Gamma_1\Gamma_2]
\la0|\bar{\psi}\psi|0\ra^2\ ,
\eqe
where $N=4\times N_f\times N_c$, and the $\Gamma_i\ ,\ (i=1,2)$, are 
direct products of Dirac, flavor, and color
matrices, one can easily verify \cite{SVZ} that 
\eab
\label{Pexp}
\la0|P_2|0\ra(\mu)&=&\frac{16}{3}\la0|P_1|0\ra(\mu)\ ,\nonumber\\ 
\la0|P_3|0\ra(\mu)&=&\la0|P_4|0\ra(\mu)=0\ ,\nonumber\\ 
 \la0|P_5|0\ra(\mu)&=&\frac{16}{3}\la0|P_6|0\ra(\mu) \ .
\eae
With \eqs {O6dec}), (\ref{mixla}), (\ref{diagel}), and (\ref{Pexp}) we 
obtain a description for the thermal 
vacuum contributions of scalar operators to the Gibbs averaged OPE of \eq {OPETl}). 

As already mentioned in the beginning of this subsection and following Ref.\cite{Hatsuda}
we restrict the thermal trace of \eq {thcor})  
to the contribution of
 one-particle pion states (the spectral integral over the pion energy can 
 savely be extended to infinity
 because of the strong Boltzmann suppression of the integrand). 
 We use the results of Ref. \cite{Hatsuda} which for ${\cal O}^s_{4}$ were obtained
  by appealing to the trace anomaly of the QCD energy-momentum-tensor. 
  In the chiral limit there is no 
 explicit temperature dependence of the thermal average of the 
 two-gluon operator ${\cal O}^s_4$
 due to pion contributions.  Applying the soft pion 
 theorem twice and using the vacuum saturation
 hypothesis of \eq{vacsat}), according to 
 Hatsuda et al. \cite{Hatsuda} one obtains for the pionic part  
 of the Gibbs average of the 
 operators ${\cal O}^s_{6,1}$ and ${\cal O}^s_{6,2}$ 
\eab
\label{pionHats}
& &\sum_{a=1}^3\int\ \frac{d^3p}{2|\vec p|}\ \la \pi^a(\vec p)| 
\frac{1}{2}{\cal O}^s_{6,1}+
\frac{1}{9}{\cal O}^s_{6,2}|\pi^a(\vec p)\ra\  n_B(|\vec p|/T)\nonumber\\ 
&=&\frac{128}{81}\pi\alpha_s\la\bar{q}q\ra^2
\left(1-\frac{3\ T^2}{8 f_\pi^2}\right)\ .
\eae
In the chiral limit we have $f_\pi=88$ MeV \cite{Mei},
 $n_B$ is the Bose distribution, and $q$ indicates a single light 
flavor quark field. As in Ref.\cite{Hatsuda} we only consider 
perfect vacuum saturation since for the finite temperature evaluation 
we are only interested in relative changes as compared to the $T=0$ case.

\subsection{O(3) invariant Operators}
\label{subsec:3.2}

The O(3) invariants of the OPE of \eq{OPETl}) have vanishing vacuum expectation values
\footnote{For diagonal operators, we can, by
 means of the scaling relation of \eq{scaling}), express their $T$-dependent 
vacuum average by a scaling factor times the $T=0$ vacuum average which vanishes due 
the boost invariance of the vacuum state.} 
and therefore we only have to consider their pionic matrix elements. 
In the chiral limit 
we encountered the following $O(3)$ invariant operators of mass-dimension four 
in the OPE of \eq{OPETl})
\eqb
\mbox{dim} 4:
\ \ \ {\cal O}^{O(3)}_{4,1}=\theta_{00}\ ,\ \ 
{\cal O}^{O(3)}_{4,2}=\frac{16}{3}\theta_{00}^f-\theta_{00}^g\ .
\eqe
The operators ${\cal O}^{O(3)}_{4,1}$ and ${\cal O}^{O(3)}_{4,2}$ 
are diagonal combinations with respect to the anomalous mixing matrix
 of $\theta^f_{00}$ and $\theta^g_{00}$
 \cite{Mal-M}. The pion matrix elements of the quark part 
  $N_f\theta^f_{00}$ ($N_f$ denotes the number of quark flavors considered) 
  and the gluon contribution 
 $\theta^g_{00}$ to the total energy density $\theta_{00}$ are 
 estimated to be equal and can be calculated 
 from the valence parton distribution in the pion 
 as found by Glueck et al. \cite{Glueck,Hatsuda}. 
 Thereby a low-energy ($Q^2=0.25$ GeV$^2$) valence like parton distribution fitted to 
 experimental data (direct photon production and
 Drell-Yan) is evolved at one and two loop order to the sum rule 
 scale of about $Q^2=1.0$ GeV$^2$. One obtains
\eab
&&\sum_{a=1}^3\int\ \frac{d^3p}{2|\vec p|}\ \la \pi^a(\vec p)| 
\theta^f_{00}|\pi^a(\vec p)\ra(Q)\ n_B(|\vec p|/T)=\nonumber\\ 
&&\frac{\pi^2 T^4}{120}\sum_{a=1}^3 A_2^{a(u+d)}(Q)\ ,
\eae
with 
\eqb
A_2^{a(u+d)}(1\, \mbox{GeV})=0.972\ , \ \forall a \ .
\eqe
At mass-dimension six we consider according to Ref.\cite{Hatsuda}
only the following twist two $O(3)$ operator 
\eqb
\mbox{dim} 6:
\ \ \ {\cal O}^{O(3)}_{6}=
i\left(\bar{u}\gamma_0D_0D_0D_0 u+(u\to d)\right)\ .
\eqe
Again with the model of Glueck et al. and in the chiral limit the 
result for the pionic contribution to the Gibbs
average of ${\cal O}^{O(3)}_{6}$ has been determined in Ref.\,\cite{Hatsuda} as  
\eab
&&\sum_{a=1}^3\int\ \frac{d^3p}{2|\vec p|}\ \la \pi^a(\vec p)| 
{\cal O}^{O(3)}_{6}|\pi^a(\vec p)\ra(Q) \ n_B(|\vec p|/T)=\nonumber\\ 
&&-\frac{2}{5}
\left(\frac{\pi^4 T^6}{63}\sum_{a=1}^3 A_4^{a(u+d)}(Q)\right)\ ,
\eae
where 
\eqb
A_4^{a(u+d)}(1\, \mbox{GeV})=0.255\ ,\ \forall a \ .
\eqe
The pion matrix elements of the operators ${\cal O}^{O(3)}_{4,1}$, 
${\cal O}^{O(3)}_{4,2}$, and  ${O(3)}_{6}$ are of order $T^4$ or 
higher which led the authors of 
Ref. [2] to omit them in the sum rule since otherwise higher orders in 
$T$ would also have to be calculated for
the scalar contributions which a priori are of order $T^2$. 
We do not agree with this power counting argument. 
The fundamental approximation for the inclusion of pions in the 
Gibbs average is that of a dilute pion gas. Higher orders in $T$ 
would come in by considering interactions
between pions as it was shown in Refs.\,\cite{Gasser,Gerber}. 
Since the pionic matrix elements of the above nonscalar operators 
happen to be of order $T^4$ and higher already 
when interactions are switched off the ommission of these 
operators would have the effect of 
artificially introducing pionic interactions. 
Hence we do not omit the nonscalar operators in the sum rule.      

We now have all the ingredients to the Gibbs averaged OPE of \eq{OPETl}).

\section{Sum Rule}
\label{sec:4}

Performing a 
Borel transformation of the OPE of Eq.\ (\ref{OPETl}), 
we obtain in the chiral limit 
the thermal sum rule for the
invariant $T_l$:
\eab
\label{srTl}
&&T_l(M)=\frac{1}{8\pi^2}(1+\frac{\alpha_s(\mu)}{\pi})+
\frac{1}{M^4}\left\{\la\frac{\alpha_s}{24 \pi}
F^a_{\mu\nu}{F^{\mu\nu}}^a\ra_T(\mu)+\right.\nonumber\\ 
& &\left.\frac{2}{11}\left[\la \theta_{00}\ra_T(\mu)+
\left(\frac{\alpha_s(\mu^2)}
{\alpha_s(M^2)}\right)^{-\delta/b}\la\frac{16}{3}
\theta_{00}^f-\theta_{00}^g\ra_T(\mu)\right]\right\}-\nonumber\\
& &\frac{1}{M^6}\left(\frac{1}{12}\la-\frac{5}{3}P_4
+\frac{32}{9}P_3-2\ P_5\ra_T(M)+\right.\nonumber\\ 
& &\frac{1}{18}\la P_4+2\ P_2\ra_T(M)+
\left.\frac{4\pi i}{3}\la\bar{q}\gamma_0D_0D_0D_0 q\ra_T (\mu)\right)\ ,
\eae
where the dispersion integral for $T_l(M)$ 
appearing on the left-hand side of \eq{srTl}) 
is given by Eqs.\ (\ref{cont}) and (\ref{emp}). 
The scalar mass-dimension six operators $P_1, P_3, P_4, P_6$ renormalized 
at $M$ can be expressed by the 
operators $P_2$ and $P_5$ renormalized at 
$\mu=1$ GeV with the help of \eqs{mixla}) and (\ref{Pexp}).

In order to eliminate the 
coupling $f_\rho$ we solve Eq.\ (\ref{srTl}) for the $\rho$\,-meson term
\dmb
R\equiv m_\rho^2\e^{-m_\rho^2/M^2}\ ,
\dme
substitute $\tau\equiv 1/M^2$, and 
perform the logarithmic derivative of $R$ \cite{Dosch,Bert}
 to obtain the ratio of moments
\eqb
\label{mrhoq}
 m_\rho^2(T,s_0,\tau)=-\frac{\partial}{\partial \tau} \mbox{log}\ R(T,s_0,\tau) \ 
\eqe
which we will consider.

\section{Numerical evaluation}
\label{sec:5}

In this section we discuss the numerical evaluation of 
the sum rule of Eq.\ (\ref{mrhoq}). 
Thereby, we employ the following values for the 
one-flavor quark and the gluon condensate \cite{Leu1}:
\eab
\label{condval}
\la 0|\bar{q}{q}|0\ra\ (\mu=1\ \mbox{GeV})&=&-(250 \ \mbox{MeV})^3\ ,  \nonumber\\  
\ \la 0|\frac{\alpha}{\pi} 
F^a_{\mu\nu}F_a^{\mu\nu}|0\ra\ (\mu=1\ \mbox{GeV})&=&0.012\ \mbox{GeV}^4\ .
\eae
As a result of vacuum sum rule calculations in a variety 
of channels the central value for the gluon condensate 
is actually believed to be twice as high as the value stated in \eq{condval}) 
with an error of about 50 \%. The central value for the quark 
condensate indicated here is rather up to date and 
exhibits a smaller error of about 30 \% \cite{Sommerschule}. However, the higher values for the gluon condensate have 
been obtained using numerical correction factors $\kappa$ 
of about $\kappa=2$ which multiply the four-quark condensate derived 
from the vacuum saturation hypothesis \cite{SVZ}. For exact vacuum saturation ($\kappa=1$) 
the values of \eq{condval}) for the quark and 
gluon condensates nicely reproduce the $\rho$\,-meson 
mass in the vacuum ($m_\rho(T=0)\approx 760$ MeV). Since we are only interested in relative changes 
of the spectral parameters and condensates for $T>0$ 
as compared to the zero temperature case 
we will stick to the values of \eq{condval}) \cite{Hatsuda}.\\  
\noindent Our strategy in evaluating the sum rule of \eq{mrhoq}) is as follows: \\ 
Since $\tau$ is not a 
direct physical observable, it is chosen
as the stationary point 
$\tau_s$ of $m_\rho^2(0,s_0(0),\tau)$ for a given $s_0$ at $T=0$. 

For a value of $s_0(0)=1.5$ GeV$^2$ an additional contribution 
to the spectral function due to $a_1-\pi$ production is regarded 
as a part of the pQCD continuum. This is reasonable as 
long as the Borel parameter $\tau$ is larger than 1 GeV$^{-2}$ 
since structures stemming from the $a_1-\pi$ production 
and possible radial excitations of the $\rho$\,-meson are then 
sufficiently suppressed in the spectral integral \cite{Narison}. 

There is a pronounced minimum for a numerically obtained 
value of $M^2_s\equiv 1/\tau_s\approx0.73$ GeV$^2$ 
corresponding to a $\rho$\,-meson mass of 745 MeV in the chiral limit considered here. 

For finite temperatures $T$ 
we determine the value of $s_0(T)$ from the 
stationary point of $m_\rho^2(T,s_0(T),\tau)$ 
at the same value $\tau=\tau_s$ as in the zero temperature case. 
Performing this calculation at a 
sufficient number of $T$ points yields 
the temperature evolution of $m_\rho^2$, $s_0$ and,
 with the parametrization of \eq{scaling}),
  also the temperature 
 dependence of the gluon condensate. The usefulness of the procedure to 
 determine the $T$-evolution for a fixed $\tau_s$ was checked by applying this 
 technique to the 
 $\rho^0$ sum rule given in Eq.\ (4.4) of Ref. \cite{Hatsuda} ($s_0(0)=1.5$ GeV$^2$). 
 Our method yields the same 
 results as found in Ref. \cite{Hatsuda}, where a more 
 elaborate evaluation analysis (averaging over a $T$-dependent Borel window) was used. 
 
We consider the following two cases: $(a)$ naive rescaling 
\footnote{Naive rescaling 
means, that, according to \eq{diagel}), the anomalous renormalization group 
 rescaling of the operators (powers of logarithms) are suppressed, 
 leaving only the factor $\lambda^6$ at mass-dimension six.} and $(b)$ 
 renormalization group rescaling 
 when including terms up to quadratic order in 
 $\lambda^2-1$ in the expansion of $D_i$ given in \eq{diagel}). 
 
The temperature evolution is determined in steps of $\Delta T=3$ MeV. 
For $s_0(0)=1.5$ GeV$^2$ and in the case $(b)$ the
 $\rho$\,-meson mass exhibits an increase 
of about 17.5\% from its zero temperature value at a `critical 
temperature' of $T_c= 157$ MeV which is 
in contrast to the result of Ref. \cite{Hatsuda}. 
Thereby, $T_c$ is obtained by 
demanding the following: If a variation of $s_0(T)$ with respect 
to $s_0(T-\Delta T)$ is larger 
than 0.3 GeV$^2$ to produce the 
 minimum of the Borel curve for $m_\rho$ at $\tau_s$, 
 we set $T=T_c$. This critical value merely indicates the breakdown of 
 our sum rule analysis and may not coincide with the 
 critical temperature of a deconfinement phase transition. 
 There is no strong dependence of $T_c$ on the 
 initial value of $s_0$ for the case $(b)$ -- $T_c(s_0(0)=1.2$ GeV$^2$)= 157 MeV, 
 $T_c(s_0(0)=1.5$ GeV$^2$)=157 MeV , 
and $T_c(s_0(0)=1.8$ GeV$^2$)=163 MeV. Fig. 1 shows the results for $m_\rho$ as a function of the 
temperature, where for $s_0(0)=1.2$ GeV$^2$ and 
$s_0(0)=1.8$ GeV$^2$ only the case $(b)$ has been considered. 
There is practically no difference in the evolution of $m_\rho$ 
when including the quadratic terms in 
$\lambda^2-1$ for the $D_i$ of \eq{diagel}) as compared to the 
linear and to the zeroth order approximation. 
The deviation at $T=160$ MeV is then at most 5 MeV. 

Fig. 2 indicates the temperature dependence of the pQCD threshold 
$s_0(T)$ for the three initial values and the case $(b)$. 
In addition, we show the
case $(a)$ choosing $s_0(0)=1.5$ GeV$^2$. Again, there is 
practically no difference between the results 
for truncations of $D_i$ in zeroth, linear, and quadratic order. Hence 
the logarithmic corrections in the Borel parameter $\tau$ seem to have a 
much greater effect on the $T$-evolution of $m_\rho$ and $s_0$ 
than the logarithmic corrections in $\lambda^2$. For $s_0(0)=1.5$ GeV$^2$ the 
 behavior in the case $(b)$ is qualitatively 
similar to that found in Ref. \cite{Hatsuda}. However, the influence of
 $s_0(T)$ on the sum rule is quite different in our work. 
 Besides the truncation of the hadronic part of the spectral function, 
 $s_0(T)$ also scales the vacuum part of the Gibbs
 averages in the OPE. In the case $(a)$ we obtain 
 an increase of $s_0(T)$ and hence a rise of the gluon condensate with 
 increasing temperature which is in contradiction to the results obtained 
 on the lattice and by effective meson models \cite{Mil,Mei,Heinz}. 

In Fig. 3 we finally show the temperature evolution of 
the gluon condensate normalized to 
its $T=0$ value which 
 due to the scaling relation of \eq{scaling}), 
 the invariance under a change of the renormalization point, and 
 the explicit $T$-independence in 
 the chiral limit (according to \eq{scaling})) is proportional to $s_0(T)^2$. 
 We indicate the same combinations of initial values $s_0(0)$ and cases
 $(a)$ and $(b)$ as for the $T$-evolution of $s_0$. Fig. 3 indicates a 
 universality of the $T$ evolution in a sense, 
 that for different values of $s_0(0)$ almost
  identical results are obtained.

\section{Summary and Discussion}
\label{sec:6}

The main concern of this paper was the investigation of the 
consequences of a temperature dependent vacuum for 
the $T$-evolution of the $\rho$\,-meson mass 
and the gluon condensate in the chiral limit. 
Thereby, we used the method of thermal QCD sum rules. 
The results obtained in Refs. \cite{Hatsuda,Mal-M} for the thermal 
Operator Product Expansion containing also O(3) invariant contributions
 and for the thermal $\rho^0$ spectral function 
have been combined in our calculations. Following Ref. 
\cite{Hatsuda}, the Gibbs averages of local, gauge invariant 
operators contributing to the thermal OPE of the 
$\rho^0$ current-current correlator were saturated 
by vacuum and one-pion matrix elements. 
For lack of better knowledge as far as the $T$-dependence of pionic 
twist two matrix elements is concerned we had to work with the 
$T=0$ parton distributions of Ref.\,\cite{Glueck}. 
Consistency then demanded the use of a $T$-independent $f_\pi$, 
although the $T$-dependence of this quantity 
is known for low temperatures from 
chiral perturbation theory \cite{Gerber}.    

In contrast to 
previous sum rule calculations we suggested to 
account for a temperature dependence of the vacuum part of the Gibbs average. 
This was motivated by the
observation, that the gluon 
condensate remains practically $T$-independent up to 
temperatures of 200 MeV in the sum rule analysis of
 Ref. \cite{Hatsuda}, while a lattice measurement 
yields a drastic decrease at a critical 
temperature $T_c$ of about $T_c\approx 140$ MeV \cite{Mil}. 
To resolve this contradiction, we derived a 
scaling relation for the thermal vacuum average 
of scalar operators with the $T$-dependent spectral pQCD
continuum threshold $s_0$. Thereby, the chiral limit, temperatures 
considerably smaller than the QCD scale $\Lambda$, and a 
renormalization point of the order of 1 GeV were assumed. 
The implementation of the above scaling relation is trivial 
for renormalization group invariant 
operators and becomes rather 
involved for operators which mix and scale anomalously 
under a change of the renormalization point. 
On the spectral side we used a narrow width 
approximation for the $\rho^0$ resonance since including a 
finite width $\Gamma$ leads to ambiguities 
in the determination of $m_\rho$ and $\Gamma$ as was shown in Ref.\ \cite{Leu1} 
for the case of finite density. One would expect, 
that this is also true for finite temperatures. On the other hand, 
the effect of the scaling of the vacuum 
averages can only be isolated if one compares 
the results with previous calculations also using 
the narrow width approximation. In analogy to a previous sum rule 
calculation in the $\rho^0$\,-channel (see Ref.\ \cite{Mal-M}) we employed a thermal pion 
continuum for the hadronic part, while the high energy tail 
was approximated by a $T$-independent pQCD continuum.  
 
As compared to the vacuum case 
our sum rule calculations indicate a rise of the 
$\rho$\,-meson mass (17.5\% at $T_c$) 
and a decrease of $s_0$ and of the gluon condensate with temperature.  
Thereby, the critical temperature $T_c$, 
where the sum rule analysis breaks down, 
is $T_c\approx 160$ MeV. This "critical" temperature is rather close 
to the QCD scale $\Lambda$, and therefore we do not expect the 
scaling relation of Eq.\ (\ref{scaling}) to be a 
good approximation for the thermal vacuum average of scalar operators. 
The increase of the $\rho$\,-meson 
mass contradicts the sum
rule results obtained in 
Refs. \cite{BS,Hatsuda,Dosch}, 
where a monotonic decrease of this quantity is obtained. 
In the difference sum rule analysis of 
Ref. \cite{Mal-M} the $\rho$\,-meson mass was also found to 
increase slightly ($\Delta m_\rho\approx$ 5 MeV) up to a temperature of about 
125 MeV while decreasing thereafter. 
There are indications for a slight rise of 
the $\rho$\,-meson mass up to $T=150$ MeV ($\Delta m_\rho\approx 30$ MeV at $T$=150 MeV) 
from a microscopical 
calculation of the spectral 
function in the $\rho^0$ channel using 
an effective $\rho-\pi$ Lagrangian \cite{Gale}. 
We emphasize again, that the critical 
temperature of our calculation does only 
indicate the breakdown of the 
sum rule analysis based on a scaling relation 
for the scalar condensates and may 
not coincide with the critical 
value for a deconfinement phase transition. 
It is interesting to note, that a calculation in the Nambu-Jona-Lasinio model 
yields at $T\approx 160$ MeV and already at nuclear 
saturation density a relative increase 
of the dynamically generated constituent light-quark mass 
which is of the same order 
as that of the $\rho$\,-meson mass 
in our calculation \cite{Tsu1,Tsu}. In Ref. \cite{Dey}, 
where by means of PCAC, current algebra, and the 
LSZ reduction formula a mixing of the vector with the axial
vector correlator for space-like $q^2$ was found 
for small temperatures (no OPE was used), a difference 
sum rule analysis 
yields an increase of the $\rho$\,-meson 
mass and a decrease of the 
mass of the $a_1$\,-meson with temperature. 
This was interpreted as 
the process of chiral symmetry restoration. 
The analysis of Ref.\,\cite{Dey} is meaningful for small temperatures ($T<f_\pi$) 
and for a $T$ independent $f_\pi$ consistent with the recent 
lattice simulation of Ref.\,\cite{BoydGupta} 
indicating no sign of a $T$ 
dependence up to $T\approx T_c$ of the chiral transition. 
Our result for the $T$-evolution 
of the $\rho$\,-meson mass is consistent 
with the one obtained in the analysis of Ref.\,\cite{Dey}. 
In addition, qualitative 
arguments based on the instanton model 
\cite{Shur2} imply a cancellation of the 
effects of a decreasing 
constituent quark mass and a lowering 
of the instanton mediated attraction between 
constituent quarks for 
the $T$-evolution of the $\rho$\,-meson mass. 

 Depending on $s_0(0)$, 
our result for the gluon condensate exhibits a 25--30\% 
decrease at $T\approx 160$ MeV which is 
compatible with the lattice data for 
two quark flavor QCD of Ref. \cite{Mil}. 
The analysis showed, that the results are sensitive 
to the anomalous scaling of mass-dimension
six operators under a change of 
the renormalization point (see Fig. 3). 
As Fig. 3 indicates the $T$-evolution of the gluon condensate 
is universal in a sense, that it is 
practically independent of 
the initial value $s_0(0)$. 

There are still open questions concerning the inclusion 
of nonscalar, mixed operators at mass-dimension
six in the thermal OPE. So far we must 
be content with the hope, that in the future 
their pionic matrix elements can 
be estimated from deep inelastic 
lepton scattering off the pion target \cite{Hatsuda}.  

To conclude, our analysis indicates, that in the Gibbs average 
the $T$-dependence of the 
vacuum matrix elements of scalar operators, as 
introduced via a scaling relation in $s_0$, 
produces a sizable decrease of the gluon 
condensate and a moderate increase of the $\rho$\,-meson
mass up to temperatures of 
about 160 MeV, where the analysis breaks down (
and the confidence in a good accuracy of 
the approximations made is not high anymore).   

\begin{acknowledgement}

The work of R.H. was supported by the Graduiertenkolleg
``Struktur und Wechselwirkung von Hadronen und Kernen''(DFG Mu705/3), and 
T.G. and A.F. were supported by the BMBF (contract no. 06 T\"u
887). One of us (R.H.) would like to express his gratitude towards Markus Eidemueller 
for reading the manuscript and suggesting some improvements. 

\end{acknowledgement}

\appendix

\section{}
\label{sec:7}

Under the premises that $T$ be 
considerably less than the fundamental 
QCD scale $\Lambda$, and that the renormalization scale $Q_0$ is taken of the 
order of 1 GeV we may argue in favor for the assumption of an implicit 
dependence of the scalar condensates on $T$ via $s_0(T)$ in the chiral limit.   

At $T=0$ the most fundamental description of 
the QCD vacuum structure by 
the appearance of nonvanishing vacuum averages of local, scalar, 
and gauge invariant operators may in the chiral limit only involve 
the scale $\Lambda$ and a renormalization scale $Q_0$. 
With the number of independent parameters kept fixed, 
we hypothesize that there is a linear relation 
between $\Lambda$ and $\sqrt{s_0(0)}$\footnote{There is then no other, independent scale 
to allow for a different behavior since $s_0$ 
cannot depend on the external variable $Q_0$.}. In this picture the constant of proportionality 
expresses this channel dependence. 
In the following we will argue that for $T>0$ 
and provided that $T$ is considerably smaller than $\Lambda$ it is 
approximately possible to define a $T$-dependent 
fundamental scale $\Lambda_T$ such that 
the thermal vacuum averages of scalar operators have the 
same functional dependence on $\Lambda_T$ and $Q_0$ 
as in the case of $T=0$, where they depend on $\Lambda$ and $Q_0$. Then there should 
be the identical linear relation 
between $\Lambda_T$ and $\sqrt{s_0(T)}$ 
as between $\Lambda$ and $\sqrt{s_0(0)}$ at $T=0$.\\   
For the thermal vacuum average of ${\cal O}$ renormalized at $Q_0$ we may write
\eqb
\la0|{\cal O}|0\ra_T(Q_0)=f(T/\Lambda,T/Q_0,\Lambda/Q_0)\times Q_0^d\ ,
\eqe
where $f$ denotes a dimensionless function of its dimensionless arguments, and $d$ indicates
the mass dimension of ${\cal O}$. We define $\Lambda_T$ by demanding
\eqb
\label{dem}
f(0,0,\Lambda_T/Q_0)\stackrel{!}=f(T/\Lambda,T/Q_0,\Lambda/Q_0)\ .
\eqe
Solving Eq.\ (\ref{dem}) for $\Lambda_T$ yields
\eqb
\Lambda_T=\Lambda_T(\Lambda,Q_0,T)=\tilde{\Lambda}_T(T/\Lambda,T/Q_0,\Lambda/Q_0)\times \Lambda\ ,
\eqe
where $\tilde{\Lambda}_T$ is a dimensionless function of its dimensionless arguments. 
Since in our applications the renormalization scale $Q_0$ is of the order of 1 GeV and 
with $T$ considerably less than $\Lambda$, 
we can neglect the dependence of $\tilde{\Lambda}_T$ on $T/Q_0$ 
\eqb
\Lambda_T(\Lambda,Q_0,T)\approx \bar{\Lambda}_T(T/\Lambda,\Lambda/Q_0)\times \Lambda\ , 
\eqe
where $\bar{\Lambda}_T$ again is a 
dimensionless function of its dimensionless arguments.  
Let us now expand $\bar{\Lambda}_T$ in powers of $\Lambda/Q_0$
\eqb
\label{expLT}
\Lambda_T=\Lambda\times(c_0(T/\Lambda)+c_1(T/\Lambda)\Lambda/Q_0+\cdots)\ .
\eqe
Since $\Lambda_{T=0}\stackrel{!}=\Lambda$ it follows that
\eqb
c_0(0)=1\ ,\ \ \ c_i(0)=0\ ,\ \ \ \ (i\ge 1)\ .
\eqe
Then the expansion of Eq.\ (\ref{expLT}) should be well 
approximated by its first term if $T$ is considerably less than $\Lambda$, namely by 
\eqb
\Lambda_T\approx \Lambda\times c_0(T/\Lambda)\ .
\eqe
Hence we obtain a well defined (since $Q_0$-independent) $T$-dependent scale $\Lambda_T$ which 
yields the same functional dependence of the thermal vacuum average 
on $\Lambda_T$ and $Q_0$ as that of the $T=0$ vacuum average on $\Lambda$ and $Q_0$.

\section{}
\label{sec:8}

Here we derive the decomposition of \eq{O6dec}). For the operator ${\cal O}^s_{6,2}$ 
one simply uses
\dmb
\bar{q}\gamma_\alpha t^a q=\bar{q}_L\gamma_\alpha t^a q_L+\bar{q}_R\gamma_\alpha t^a q_R
\dme
to obtain
\eqb
{\cal O}^s_{6,2}=\pi\alpha_s\left(P_4+2\ P_2\right)\ .
\eqe
Using
\dmb
\bar{q}\gamma_\alpha\gamma_5 t^a q=\bar{q}_L\gamma_\alpha t^a q_L-
\bar{q}_R\gamma_\alpha t^a q_R\ ,
\dme
yields the following decomposition for ${\cal O}^s_{6,1}$ 
\eab
{\cal O}^s_{6,1}&=&\pi\alpha_s\left[(\bar{\psi}_L\gamma_\alpha t^a\tau^3\psi_L)^2-\right.\nonumber\\ 
& &\left.2\bar{\psi}_L\gamma_\alpha t^a\tau^3\psi_L
\bar{\psi}_R\gamma_\alpha t^a\tau^3\psi_R+\right.\nonumber\\ 
& &\left.(\bar{\psi}_R\gamma_\alpha t^a\tau^3\psi_R)^2\right]\ .
\eae
We are only interested in the flavor singlet part of the right-hand side, 
for which we also write ${\cal O}^s_{6,1}$ 
\eab
{\cal O}^s_{6,1}&=&\frac{1}{3}\pi\alpha_s
\left[(\bar{\psi}_L\gamma_\alpha t^a\tau^b\psi_L)^2-\right.\nonumber\\ 
& &\left.2\bar{\psi}_L\gamma_\alpha t^a\tau^b\psi_L
\bar{\psi}_R\gamma_\alpha t^a\tau^b\psi_R+\right.\nonumber\\ 
& &\left.(\bar{\psi}_R\gamma_\alpha t^a\tau^b\psi_R)^2\right]\nonumber\\ 
&=&\frac{1}{3}\pi\alpha_s\left[(\bar{\psi}_L\gamma_\alpha t^a\tau^b\psi_L)^2+\right.\nonumber\\ 
& &\left.(\bar{\psi}_R\gamma_\alpha t^a\tau^b\psi_R)^2-2 P_5\right]\ .
\eae
For the sake of brevity we only consider the following operator
\eqb
{\cal O}_{L(R)}=(\bar{\psi}_{L(R)}\gamma_\alpha t^a\tau^b\psi_{L(R)})^2\ .
\eqe
Writing out color and flavor indices (in this order), using
\dmb
\tau^c_{ij}\tau^c_{mn}=
2(\delta_{in}\delta_{jm}-\frac{1}{N}\delta_{ij}\delta_{mn})\ ,
\dme
with $N_c=3$ and $N_f=2$ one obtains
\eab
\label{sun}
&&{\cal O}_{L(R)}=\nonumber\\ 
&&4\left[\bar{\psi}_{L(R),i\mu}\gamma_\alpha \delta_{im}
\delta_{\mu\lambda}\psi_{L(R),j\nu}\bar{\psi}_{L(R),l\kappa}\gamma_\alpha 
\delta_{jl}\delta_{\nu\kappa}\psi_{L(R),m\lambda}-\right.\nonumber\\ 
& &\left.\frac{1}{2}\bar{\psi}_{L(R),i\mu}\gamma_\alpha \delta_{im}
\delta_{\mu\nu}\psi_{L(R),j\nu}\bar{\psi}_{L(R),l\kappa}\gamma_\alpha 
\delta_{jl}\delta_{\kappa\lambda}\psi_{L(R),m\lambda}-\right.\nonumber\\ 
& &\left.\frac{1}{3}\bar{\psi}_{L(R),i\mu}\gamma_\alpha \delta_{ij}
\delta_{\mu\lambda}\psi_{L(R),j\nu}\bar{\psi}_{L(R),l\kappa}\gamma_\alpha 
\delta_{lm}\delta_{\nu\kappa}\psi_{L(R),m\lambda}+\right.\nonumber\\ 
& &\left.\frac{1}{6}\bar{\psi}_{L(R),i\mu}\gamma_\alpha \delta_{ij}
\delta_{\mu\nu}\psi_{L(R),j\nu}\bar{\psi}_{L(R),l\kappa}\gamma_\alpha 
\delta_{lm}\delta_{\kappa\lambda}\psi_{L(R),m\lambda}\right]\ .\nonumber\\ 
\eae
Applying the Fierz transformation \cite{SVZ}
\eab
&&\bar{\psi}_{L(R),1}\gamma_\alpha\psi_{L(R),2}\bar{\psi}_{L(R),3}
\gamma_\alpha\psi_{L(R),4}=\nonumber\\ 
&&\bar{\psi}_{L(R),1}\gamma_\alpha\psi_{L(R),4}\bar{\psi}_{L(R),3}
\gamma_\alpha\psi_{L(R),2}
\eae
to the first and the third line of \eq{sun}) 
and relabelling $m\leftrightarrow j$ in the second line, 
 the sum of the second and 
the third line of \eq{sun}) reads
\eqb
\label{23}
-\frac{5}{6}\bar{\psi}_{L(R),i}\gamma_\alpha 
\psi_{L(R),j}\bar{\psi}_{L(R),j}\gamma_\alpha 
\psi_{L(R),i}\ ,
\eqe
where the summation over flavor indices is implicit. This can easily be rewritten as
\eqb
-\frac{5}{12}\bar{\psi}_{L(R)}\gamma_\alpha t^a 
\psi\bar{\psi}_{L(R)}\gamma_\alpha t^a
\psi_{L(R)}-\frac{5}{18}\bar{\psi}_{L(R)}\gamma_\alpha 
\psi\bar{\psi}_{L(R)}\gamma_\alpha 
\psi_{L(R)}\ .
\eqe
Putting everything together we finally obtain 
\eqb
{\cal O}^s_{6,1}=
\frac{1}{3}\pi\alpha_s\left(-\frac{5}{3}P_4+\frac{32}{9}P_3-2P_5\right)\ .
\eqe

\noindent
\newpage \vspace*{1cm}
\noindent Figure 1:\ {Temperature evolution of the $\rho$\,-meson mass. 
The solid lines correspond to the case $(b)$ for 
$s_0(0)=1.2$ GeV$^2$ ($m_\rho(T=0)=694$ MeV), $s_0(0)=1.5$ GeV$^2$ ($m_\rho(T=0)=745$ MeV), 
and $s_0(0)=1.8$ GeV$^2$ ($m_\rho(T=0)=776$ MeV). For $s_0(0)=1.5$ GeV$^2$ 
the case $(a)$ is associated 
with a dashed line.}\vspace{1cm}\\

\noindent Figure 2:\ {Temperature evolution of the pQCD threshold $s_0$. 
The solid lines correspond to the case $(b)$. For $s_0(0)=1.5$ GeV$^2$ 
the case $(a)$ is associated 
with the dashed line.}\vspace{1cm}\\ 

\noindent Figure 3:\ {Temperature evolution of the gluon 
condensate normalized to its zero temperature value. 
For the case $(b)$ we take $s_0(0)=1.2$ GeV$^2$, $s_0(0)=1.5$ GeV$^2$, 
and $s_0(0)=1.8$ GeV$^2$ corresponding to dot-dashed, solid, 
and long dashed lines, respectively. 
For $s_0(0)=1.5$ GeV$^2$ the case $(a)$ is associated 
with a dotted line.}

\newpage

\begin{figure*}[h]
\label{1}
\vskip 23.5cm 
\includegraphics{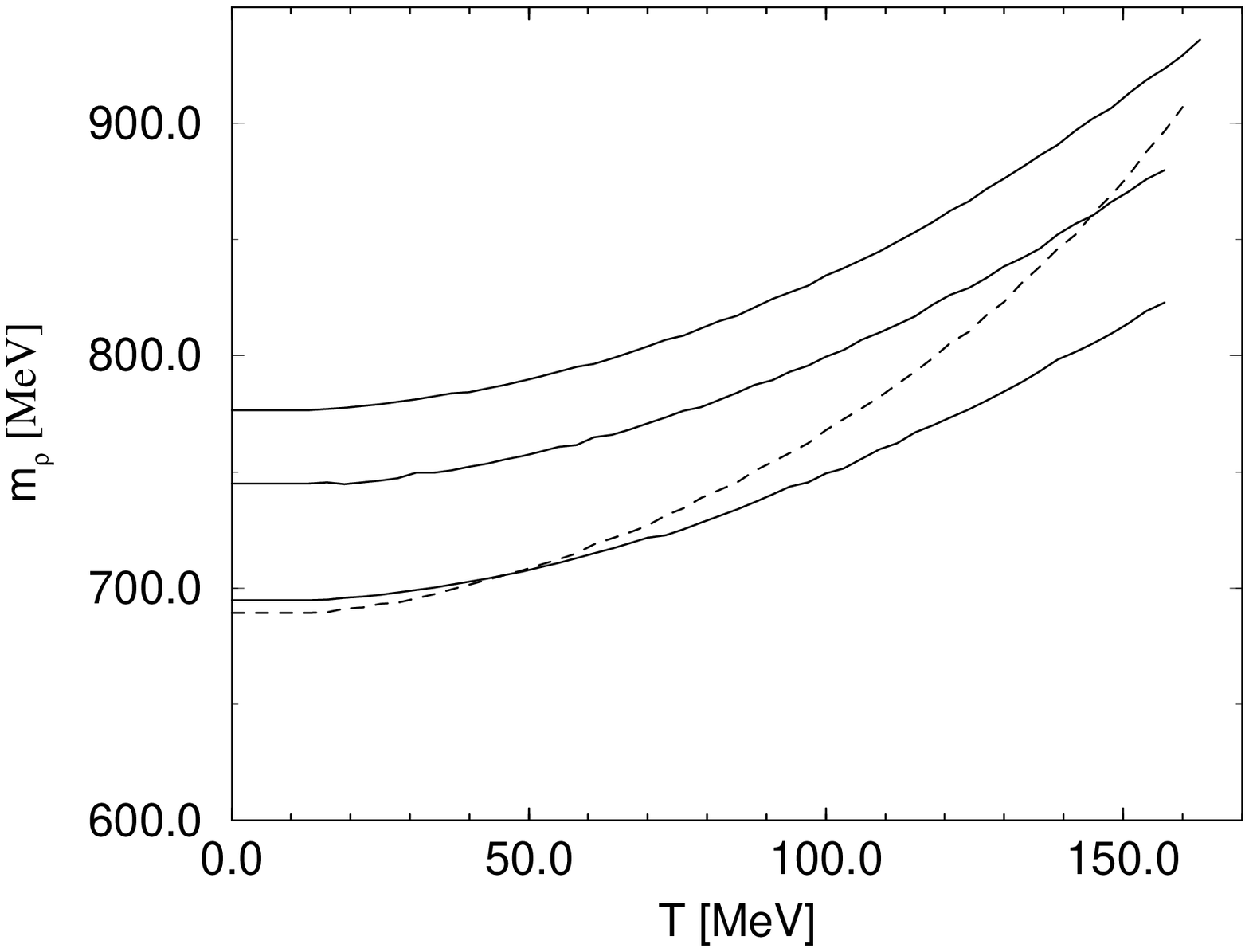}
	 \caption{} 
\end{figure*}

\newpage

\begin{figure*}[h]
\label{2}
\vskip 23.5cm 
\includegraphics{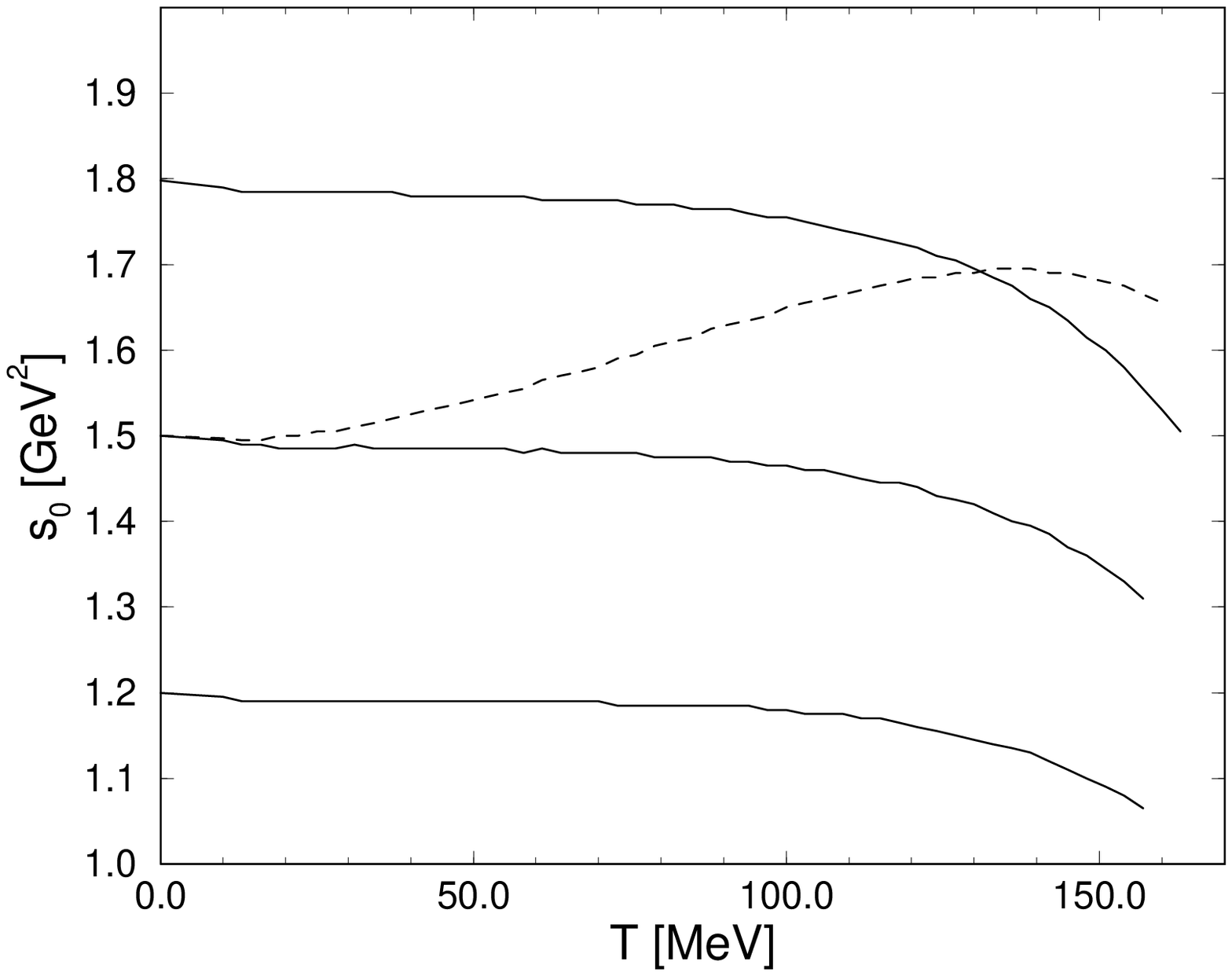}
	 \caption{} 
\end{figure*}

\newpage
  
\begin{figure*}[h]
\label{3}
\vskip 23.5cm 
\includegraphics{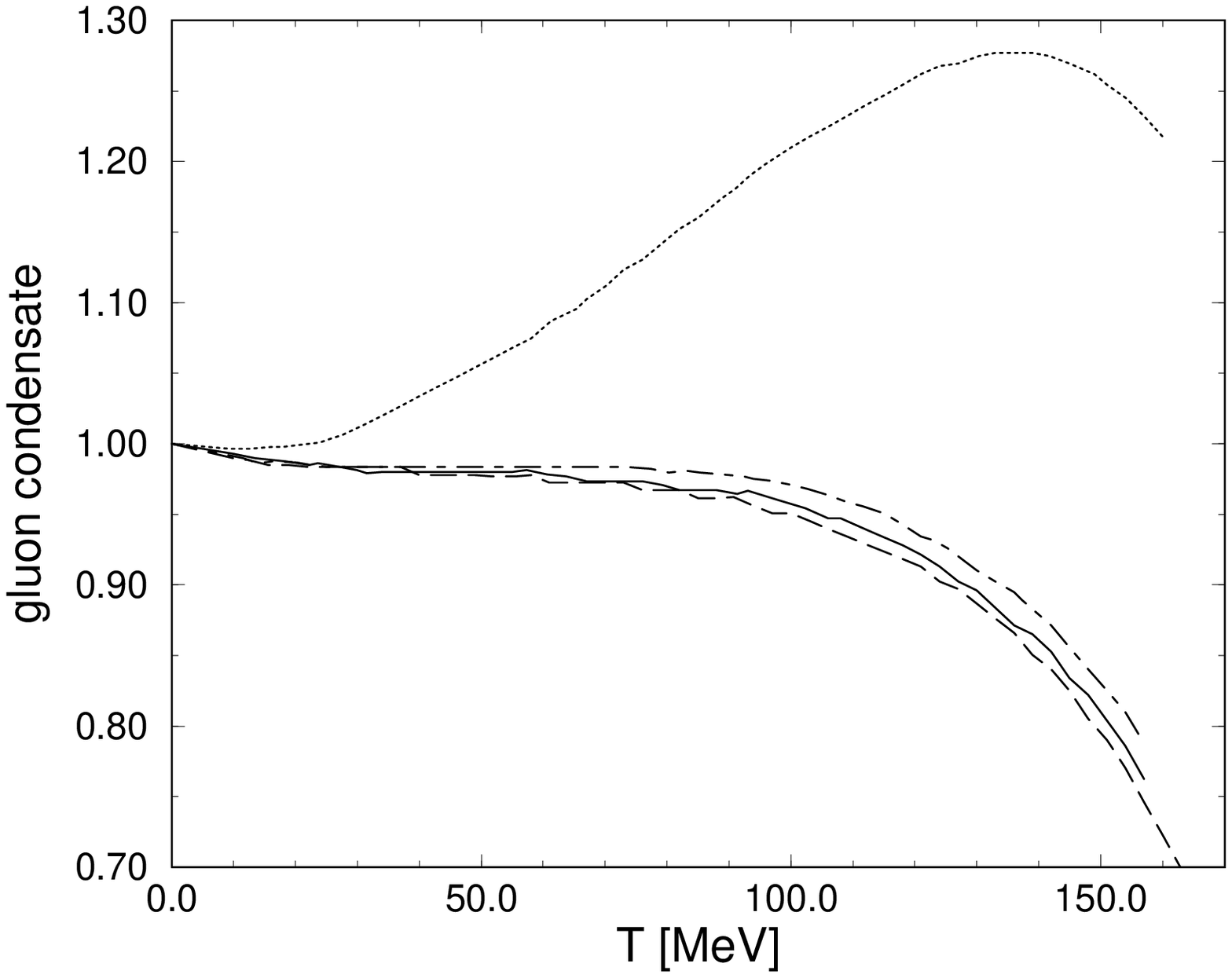}
\caption{} 
\end{figure*}

\newpage

\end{document}